\documentclass[prb,twocolumn,amsmath,amssymb,superscriptaddress]{revtex4}
\usepackage[draft]{graphicx}
\usepackage{bm}

\graphicspath{{figures/}}
\usepackage[version=4]{mhchem}

\usepackage[dvipsnames]{xcolor} 
\usepackage[colorlinks=true, citecolor = blue]{hyperref}

\newcommand{\im}{{\rm i}}
\usepackage[normalem]{ulem}

\begin{document}

\title{Effects of staggered Dzyaloshinskii-Moriya interactions in a quasi-two-dimensional Shastry-Sutherland model}

\author{Tianqi Chen}
\affiliation{Science and Math Cluster, Singapore University of Technology and Design, 8 Somapah Road, 487372 Singapore}

\author{Chu Guo}
\affiliation{Quantum Intelligence Lab (QI-Lab), Supremacy Future Technologies (SFT), Guangzhou 511340, China}      

\author{Pinaki Sengupta}
\affiliation{School of Physical and Mathematical Sciences, Nanyang Technological University, 21 Nanyang Link, 637371 Singapore}

\author{Dario Poletti}
\affiliation{Science and Math Cluster, Singapore University of Technology and Design, 8 Somapah Road, 487372 Singapore}

\date{\today}

\begin{abstract}
Frustrated quantum spin systems exhibit exotic physics induced by external magnetic field with anisotropic interactions. Here, we study the effect of non-uniform Dzyaloshinskii-Moriya (DM) interactions on a quasi-two-dimensional Shastry-Sutherland lattice using a matrix product states (MPS) algorithm. 
We first recover the magnetization plateau structure present in this geometry and then we show that both interdimer and intradimer DM interactions significantly modify the plateaux. The non-number-conserving intradimer interaction smoothens the shape of the magnetization curve, while the number-conserving interdimer interaction induces different small plateaux, which are signatures of the finite size of the system. Interestingly, the interdimer DM interaction induces chirality in the system. We thus characterize these chiral phases with particular emphasis to their robustness against intradimer DM interactions.                    
\end{abstract}

\maketitle

\section{\label{sec:introduction}Introduction}

The competition between different terms in a Hamiltonian can lead to the emergence of different phases of matter. A particularly rich set of systems, from this perspective, is that of frustrated quantum magnets in which different phases, ranging from ferromagnetism to spin liquid, have been observed \cite{Diep2013}.   
The Shastry-Sutherland (SS) model is a canonical model to study the emergence of complex 
quantum phases resulting from the interplay between strong interactions, geometric frustration, and
external magnetic field. The model has attracted widespread interest for the unique sequence
of magnetization plateaux that appear in a magnetic field. The exact ground state of the SS model in
zero field is known to be a direct product of singlets on the dimer bonds\cite{ShastrySutherland1981}.
Beyond this limit, it can be studied using a variety of approximate analytical and numerical
methods \cite{MisguichGirvin2001, DorierMila2008, AbendscheinCapponi2008, ManmanaMila2011, JaimeGoulin2012, NemecSchmidt2012, MatsudaMila2013, CorbozMila2014, WietekHonecker2019}. 
Importantly, this model is relevant in the understanding of the magnetic properties of \ce{SrCu2(BO3)2}, a material which manifests magnetization plateaux which have been observed since two decades ago \cite{KageyamaUeda1999, OnizukaGoto2000, KodamaMila2002}. 

The SS model has been investigated also in configurations different from a two-dimensional (2D) system, but also for ladders with four spins per rung \cite{ManmanaMila2011, JaimeGoulin2012} and other finite geometries\cite{RichterSchulenburg1998, KogaKawakami2000, SchulenburgRichter2002, HoneckerRichter2004, OhanyanHoenecker2012, VerkholyakStrecka2013}.

These geometries reproduce some of the physics present for 2D systems, thus giving clearer insight and understanding on them (e.g. the presence of plateaux due to triplets or to triplet bound pairs \cite{ManmanaMila2011}). Interestingly, due to the advances in ultracold atoms experiments, it could be possible to realize these systems, for instance, with tweezers and Rydberg atoms \cite{BarredoBrowayes2016, EndresLukin2016}.    

In the study of \ce{SrCu2(BO3)2} it was pointed out that Dzyaloshinskii-Moriya (DM) interactions may play a role in the groundstate properties of the system \cite{RoomKageyama2004, LevyUeda2007}, motivating a more in depth study of the effect of these interactions. Moreover, since DM interaction could potentially be produced in a large range of magnitudes in ultracold atom experiments \cite{DalibardOhberg2011, GoldmanSpielman2014}, we study values of DM interactions which are beyond what could be found in solid state materials.        
In this work we study the SS model on a ladder as shown within the orange box in Fig.\ref{fig1_lattice}(a), and we focus on how DM interactions affect the magnetization plateaux. We analyze both the effect of only interdimer or only intradimer DM interactions and then consider also their combined effect. Using a matrix product state (MPS) algorithm we show that intradimer DM interactions, which are non-number-conserving, smoothen the magnetization curve, while the (number-conserving) interdimer DM interaction alone induces a staggered chirality in the triangular plaquettes between dimers. We also show that this chiral structure is robust in the presence of intradimer interactions. 

This paper is organized as follows. In Sec. \ref{sec:model} we describe the Shastry-Sutherland model. We then discuss the numerical approach used in Sec. \ref{sec:methods}. After that, in Sec. \ref{sec:results} we present our results for the magnetization curve in the presence of DM interactions and their effect on two-point correlation functions measured on each dimer, with a particular emphasis on the emergence of chirality and on the combined effect of both intradimer and interdimer interactions. Finally, in Sec. \ref{sec:conclusion} we draw our conclusions.

\section{\label{sec:model}Our model}
\begin{figure}
\includegraphics[width=\columnwidth,draft=false]{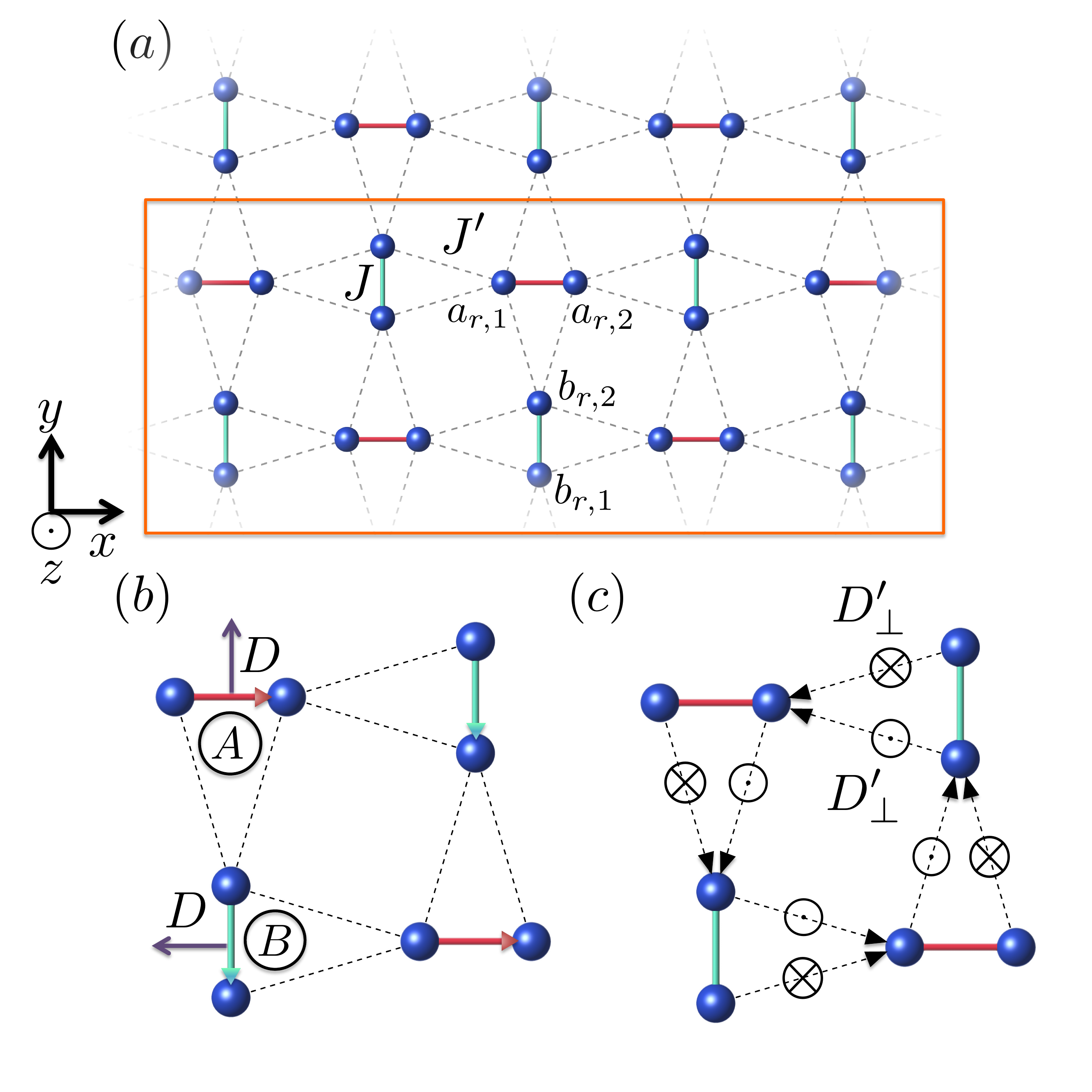}
\caption{\label{fig1_lattice}(a) Shastry-Sutherland lattice. $J$ and $J'$ labels the intradimer and interdimer bond exchange interactions, respectively. The orange square indicates a lattice with $4 \times 5=20$ sites where the periodic boundary condition is applied in the $y$ direction, and the open boundary condition is applied in the $x$ direction. In this work the lattice size we consider is $4 \times 12=48$. The $a$ and $b$ are labeling conventions for two-point correlation functions on each dimer. (b) Schematics of in-plane intradimer DM interactions. The circled letter $A$ and $B$ indicates the category of horizontal dimer (red) and vertical dimer (green), respectively. (c) Schematics of out-of-plane interdimer DM interactions} 
\end{figure}

We analyze a 2D Shastry-Sutherland (SS) ladder whose geometry is shown in the box of Fig. \ref{fig1_lattice}(a). It is composed of $L$ rungs of four spins, two in the horizontal direction, which we will refer to as the $A$ dimer, and two in the vertical direction, the $B$ dimer. 
The four spins are marked as $a_{r,1}$, $a_{r,2}$, $b_{r,1}$, and $b_{r,2}$, where $r$ labels the rung number while $1$ or $2$ differentiate the two spins within the horizontal or vertical dimer. We consider periodic boundary conditions in the $y$ direction, and up to $12$ rungs in the $x$ directions with open boundary conditions, for a total of at most $48$ spins. We study the spin-$1/2$ antiferromagnetic Heisenberg model on the SS lattice in an external magnetic field $h_z$ applied along the $z$ direction and Dzyaloshinskii-Moriya interactions. The resulting SS model is described by the Hamiltonian 
\begin{align}
\label{ssmodelhamiltonian}
\mathcal{H} &= J\sum_{\langle i,j \rangle} \bm{S}_i \bm{S}_j + J'\sum_{\langle \langle i,j \rangle \rangle} \bm{S}_i \bm{S}_j - h_z \sum_{i} S^z_{i} \nonumber\\  
& +\sum_{\langle i,j \rangle}\bm{D}_{ij} \left( \bm{S}_i \times \bm{S}_j \right) +\sum_{\langle \langle i,j \rangle \rangle}\bm{D'}_{ij} \left( \bm{S}_i \times \bm{S}_j \right)
\end{align}
where $\bm{S}_i$ is a vector of operators $\bm{S}_i=\left(S_i^x,S_i^y,S_i^z\right)=1/2\left(\sigma_i^x,\sigma_i^y,\sigma_i^z\right)$ each element being $1/2$ times a Pauli operator acting on site $i$. 
$J$ is the magnitude of the intradimer coupling (e.g., $a_{r,1}$ with $a_{r,2}$ and $b_{r,1}$ with $b_{r,2}$ ), denoted by $\langle i,j \rangle$, while $J'$ is the magnitude of the interdimer coupling (e.g., $a_{r,1}$ with $b_{r,2}$ and $b_{r,1}$ with $a_{r+1,1}$ ), denoted by $\langle \langle i,j \rangle \rangle$. Here $i$ and $j$ are labels for the spins which map the tuples $(r,1)$ or $(r,2)$ to a single integer. We denote $\mathbf{D}_{ij}$ as the intradimer and $\mathbf{D'}_{ij}$ as the interdimer DM vectors. 
Given the vectorial nature of the DM interaction, it is important to specify clearly its direction. 
In this work, we mainly follow the  convention in\cite{RomhanyiPenc2011} as shown in Figs. \ref{fig1_lattice}(b) and \ref{fig1_lattice}(c). In particular, the intradimer term $\bm{D}_{i,j}$ takes the values $\bm{D_A}=(0,D,0)$ or $\bm{D_B}=(-D,0,0)$ respectively for the horizontal and the vertical dimers when $i$ and $j$ correspond to the spins $1$ and $2$ in the dimer [see Fig. \ref{fig1_lattice}(b)] \cite{fn1}. For the interdimer interaction, {we consider the vector to take} the value $\bm{D}^{\prime}_{i,j}=(0,0,D_{\perp}^{\prime})$ with the labels $i$ and $j$ considered in the direction of the thin black arrows in Fig. \ref{fig1_lattice}(c) \cite{fn2}. {While the $\bm{D}^{\prime}_{i,j}$ vector could be pointing also in a different direction, the chosen direction would reinforce, together with $\bm{D_A}$ and $\bm{D_B}$, noncoplanar spin configurations.}         
The DM interaction thus forms a staggered pattern on the whole lattice.

\section{\label{sec:methods} Implementation of the MPS ground state search algorithm}
We numerically obtained the ground state of SS model as well as magnetization plateaux and two-point correlation functions using a ground state search algorithm implemented with MPS \cite{Schollwock2011}. Such method, together with the equivalent and earlier density matrix renormalization group \cite{White1992}, has proven to be a powerful numerical approach designed for calculating the ground state of one-dimensional quantum many-body systems. It can also be used successfully for certain $2$D systems, including frustrated spin systems, by mapping the $2$D system to a $1$D chain with interactions beyond nearest-neighbor's \cite{StoudenmireWhite2012}. 
The MPS approach is fundamentally a variational one and, for systems with a complex energy landscape, the algorithm can sometimes get trapped in local minima. Several techniques have been explored to avoid the algorithm being trapped in a local minimum, such as adding corrections to the density matrix \cite{White2005}, or perturbing the interaction Hamiltonian \cite{StoudenmireWhite2012}. In this work, we adopted similar ideas by adding random disorder on different bonds in the first sweeps. 
We also add, in the first sweeps, a small and decreasing intradimer DM interaction to remove the symmetries in the system and allow the wave function to vary more freely, similar to \cite{ManmanaMila2011}. 
In our simulations we have applied cylindrical boundary conditions to the system, i.e., periodic boundary conditions are applied in the $y$ direction and open boundary conditions are used in the $x$ direction, and we consider a system with a length of $12$ unit cells, so that the total number of sites considered is then $48$, a system size which allows one to probe various possible magnetization plateaux. We also use bond dimensions up to $M=500$.

\section{\label{sec:results}Results}    
\subsection{Magnetization plateaux and DM interactions}
\begin{figure}
\includegraphics[width=\columnwidth,draft=false]{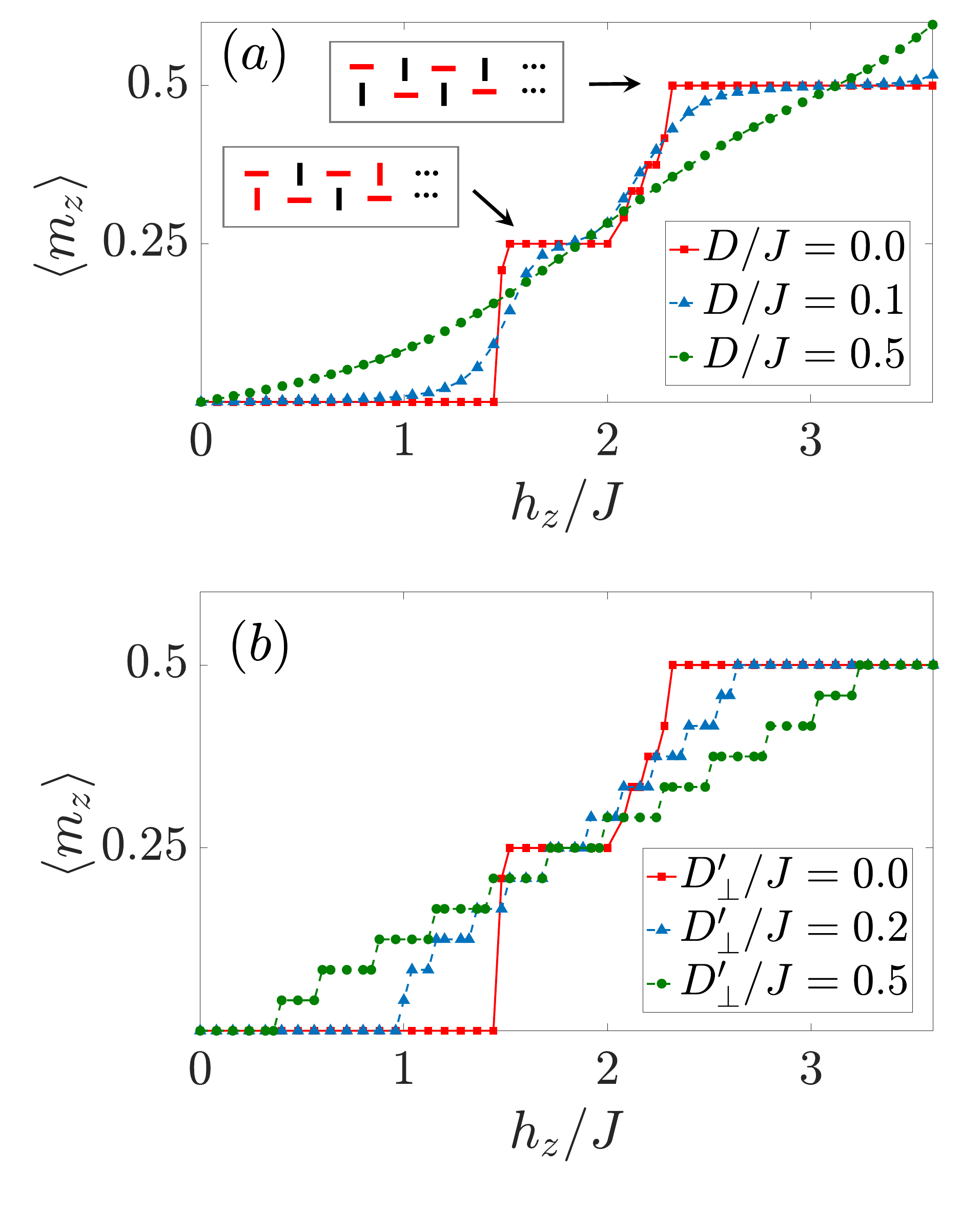}
 \caption{\label{fig2_magcurve} Magnetization curve of $48$-site SS model for $J'/J=0.5$: (a) only with intradimer DM interactions. (inset) ground state configuration within $1/4$ and $1/2$ plateaux: singlet state (red) and triplet state (black). $D/J=0.0$ (red squared solid line), $0.1$ (blue dashed line with triangles), $0.5$ (green dashed line with circles); (b) only with interdimer DM interactions. $D_{\perp}'/J=0.0$ (red squared solid line), $0.2$ (blue dashed line with triangles), $0.5$ (green dashed line with circles).} 
\end{figure}
We study the average magnetization in the ground state of the system as a function of $h_z$ using 
\begin{align}
	\langle m_z \rangle &= \sum_i \frac{\langle  \psi |  S_i^z | \psi \rangle}{4L }
\end{align}
where $|\psi \rangle$ is the ground state and $4L$ is the total number of spins. 
In the following we set $J'/J=0.5$, so that the zero-field ground state is in a strong dimer phase. 
The magnetization curves with and without DM interactions are plotted in Fig. \ref{fig2_magcurve}. 

In the absence of DM interaction, and for our choice of $J'/J$, system size, boundary conditions, and magnetic field magnitudes, {the magnetization curve presents two extended plateaux for $m_z=1/4$ and $m_z=1/2$. This is consistent with the phase diagram shown in \cite{ManmanaMila2011} in which the authors considered the same geometry, although with more rungs.\footnote{A narrow plateaux at $m_z=1/3$ reported in Ref. \cite{ManmanaMila2011} is not observed in our results. This is due to the smaller system sizes in the present study that are inadequate to resolve the narrow $1/3$ plateau.} } 

The ground state configuration on each dimer for the $1/4$ and $1/2$ plateaux are shown in the inset in Fig. \ref{fig2_magcurve}(a). For the $1/2$ plateau, a large overlap with one singlet $1/\sqrt{2}\left(|\uparrow \downarrow\rangle-|\downarrow\uparrow\rangle\right)$ and one triplet $|\uparrow\uparrow\rangle$ appear within each unit cell. For the $1/4$ plateau two unit cells are required to host four dimers of which three are close to singlets and the remaining one is mostly a triplet. These triplets are in a configuration such that they are arranged one next to another. We compare the role of intradimer and interdimer DM interactions in affecting the magnetization curves respectively in Fig. \ref{fig2_magcurve}(a) and Fig. \ref{fig2_magcurve}(b). 
As shown in Fig. \ref{fig2_magcurve}(a), for an increasing $D/J$, the plateaux become smoother until they completely disappear. This is due to the fact that the intradimer DM interaction terms on horizontal and vertical dimers are respectively $\sum_{\langle i,j \rangle}S_{z,i}S_{x,j}-S_{x,i}S_{z,j}$ and, $\sum_{\langle i,j \rangle}S_{y,i}S_{z,j}-S_{z,i}S_{y,j}$ which do not conserve the magnetization in the $z$ direction and favor spin directions other than parallel or anti-parallel to the $h_z$ field. In Fig. \ref{fig2_magcurve}(b) we show the effects of solely the interdimer DM interaction. This results in the shift and/or reduction of the $1/4$ and $1/2$ plateaux and the generation of further plateaux. However, these plateaux are due to the finite size of the system and on the fact that the interdimer DM interaction is number-conserving. {For larger values of DM interaction, the ground state magnetization changes slowly over a wide range of applied fields. In other words, the energy of the ground state changes continuously with magnetization, instead of discrete jumps. However, for a finite system, the magnetization can only increase in finite steps. The field steps used in our simulation  correspond to energy changes smaller than that between states with different magnetization, resulting in the appearance of small plateaux. For larger system size, this curve will become smooth. In a few words, as the magnetic field varies, the ground state of a different total magnetization sector becomes the overall ground state, and thus we observe many plateaux. For larger system size, this curve will become smooth}. To gain further insight into the effect of the DM interactions, in the following we will study correlations in the ground state.       

\subsection{\label{subsec:GSwithDM interactions}Ground state with DM interactions} 
\begin{figure}
\includegraphics[width=\columnwidth,draft=false]{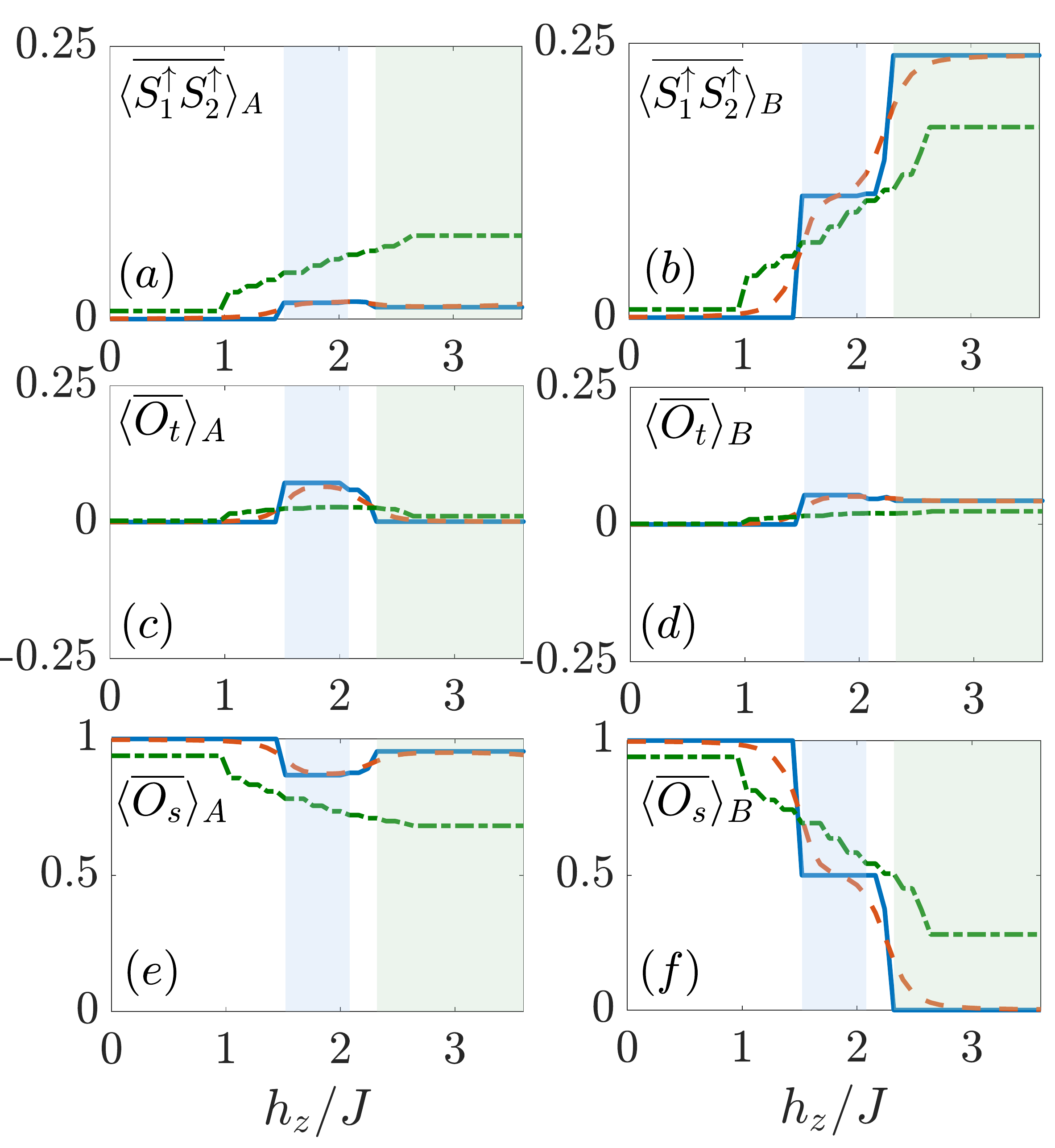}
\caption{\label{fig3} Averaged two-point correlations as a function of magnetic field $h_z$: (a),(b) $\langle \overline{S_1^{\uparrow}S_2^{\uparrow}} \rangle_{\nu}$ that probes the $|\uparrow \uparrow \rangle$ triplet; (c),(d) $\langle \overline{O_t} \rangle_{\nu}$ that probes the $1/\sqrt{2} \left( |\uparrow \downarrow \rangle + |\downarrow \uparrow \rangle\right)$ triplet; (e),(f) $\langle \overline{O_s} \rangle_{\nu}$ that probes $1/\sqrt{2} \left( |\uparrow \downarrow \rangle - |\downarrow \uparrow \rangle\right)$ singlet. The ${\nu}=A\; (B)$ in the subscript in all plots indicate the correlation functions averaged over all dimer $A\; (B)$. Results are shown for no DM interactions (blue solid line), only intradimer $D/J=0.1$ (orange dashed line), and only interdimer $D'_{\perp}/J=0.2$ (green dash-dot line) DM interactions. The two shaded areas represent the regions of parameters corresponding to the $1/4$ (blue) and $1/2$ (green) plateaux in absence of DM interactions. }
\end{figure}
In the absence of magnetic field $h_z$ and of DM interaction, the state with singlets on every site is the ground state of the system. As the magnetic field increases, the ground state can form patterns with singlet or triplet correlations on the dimers. For instance, in the $1/2$ plateaux one can expect a pattern as shown in the inset of Fig. \ref{fig2_magcurve}(a), where one type of dimer, $A$ or $B$, is close to a singlet state (we will refer to these dimers as ``singlet dimers''), while the other is close to a triplet state (we will refer to them as ``triplet dimers''). 

The boundary conditions and the search algorithm pin one of the possible translationally equivalent patterns, and hence when studying the average correlations over only the $A$ or $B$ dimers we can expect significant differences.  
For this reason, here we analyze three types of two-point correlation functions on each $A$ or $B$ dimer: $\langle \overline{S_i^{\uparrow} S_j^{\uparrow}} \rangle_{\nu}$, $\langle \overline{O_t} \rangle_{\nu}$ and, $\langle \overline{O_s}\rangle_{\nu}$, where the overline indicates an average over the ${\nu}=A$ or ${\nu}=B$ dimers in the eight middle rungs in order to reduce boundary effects. The operators $O_t$ and $O_s$ are given by
\begin{align}
\label{eq:O_tandO_s}
O_t &= 2\left(S_i^{\uparrow}S_j^{\downarrow}+S_i^{\downarrow}S_j^{\uparrow}\right) + 2 \left(S_i^+S_j^-+S_i^-S_j^+\right) \\ \nonumber 
O_s &= 2\left(S_i^{\uparrow}S_j^{\downarrow}+S_i^{\downarrow}S_j^{\uparrow}\right) - 2 \left(S_i^+S_j^-+S_i^-S_j^+\right), 
\end{align}
where we have used $S^{\pm}_i=1/2(S^x_i\pm \im S^y_i)$ and $S^{\uparrow}_i = 2 S^{+}_iS^{-}_i$. 
{The operators $O_s$ and $O_t$ project, respectively, onto the singlet, $1/\sqrt{2}\left(|\uparrow \downarrow \rangle - |\downarrow \uparrow \rangle \right)$, or triplet, $1/\sqrt{2}\left(|\uparrow \downarrow \rangle + |\downarrow \uparrow \rangle \right)$, states while the correlation$\langle \overline{S_i^{\uparrow} S_j^{\uparrow}} \rangle$ is used for probing triplet states of the type $|\uparrow \uparrow \rangle$.}
The results for triplet states of type $|\downarrow \downarrow \rangle$ are not shown in the plot as the magnitude is of $10^{-4}$ to $10^{-3}$ and thus is considered not significant compared with the other three correlators. 
  
The behavior of these correlators versus the magnetic field $h_z$ is shown in Fig. \ref{fig3} with and without both types of DM interactions. The values for no DM interactions are represented by continuous blue lines, green dashed lines have been used for interdimer DM interactions (with $D'_{\perp}/J=0.2$), and red dot-dashed lines for intradimer ones (with $D/J=0.1$). Panels (a), (c) and (e) show results for $A$ dimers, while panels (b), (d) and (f) for the $B$ ones. We stress here that in Fig. \ref{fig3} we have included two shaded areas: the blue one corresponds to the values of the magnetic field for which, in the absence of DM interactions, the $1/4$ plateau is present, while the green one signals the values of $h_z$ for the $1/2$ plateau.

When the system has no DM interactions, the blue continuous lines in Fig. \ref{fig3} show clearly that the $A$ dimers remain throughout close to a singlet state while, first, half of the $B$ dimers become close to triplets of the $|\uparrow\uparrow\rangle$, and then all of them in the $1/2$ plateau. We observe a small signal also for triplets with zero magnetization signaled by $\langle\bar{O}_t \rangle_{\nu}$. 

When considering the functional dependence of the different observables as a function of the magnetic field, we observe that, despite the presence of an intradimer interaction alone, the red-dashed curves follow, although in a smoother way, the blue-continuous curve computed in the absence of any DM interaction. The interdimer interaction alone instead results in dimers $A$, which depart more from a singlet and approach more a triplet $|\!\uparrow\uparrow\rangle$ state for larger magnetic fields, while the opposite occures for $B$ dimers.

\begin{figure*}
\includegraphics[width=2 \columnwidth,draft=false]{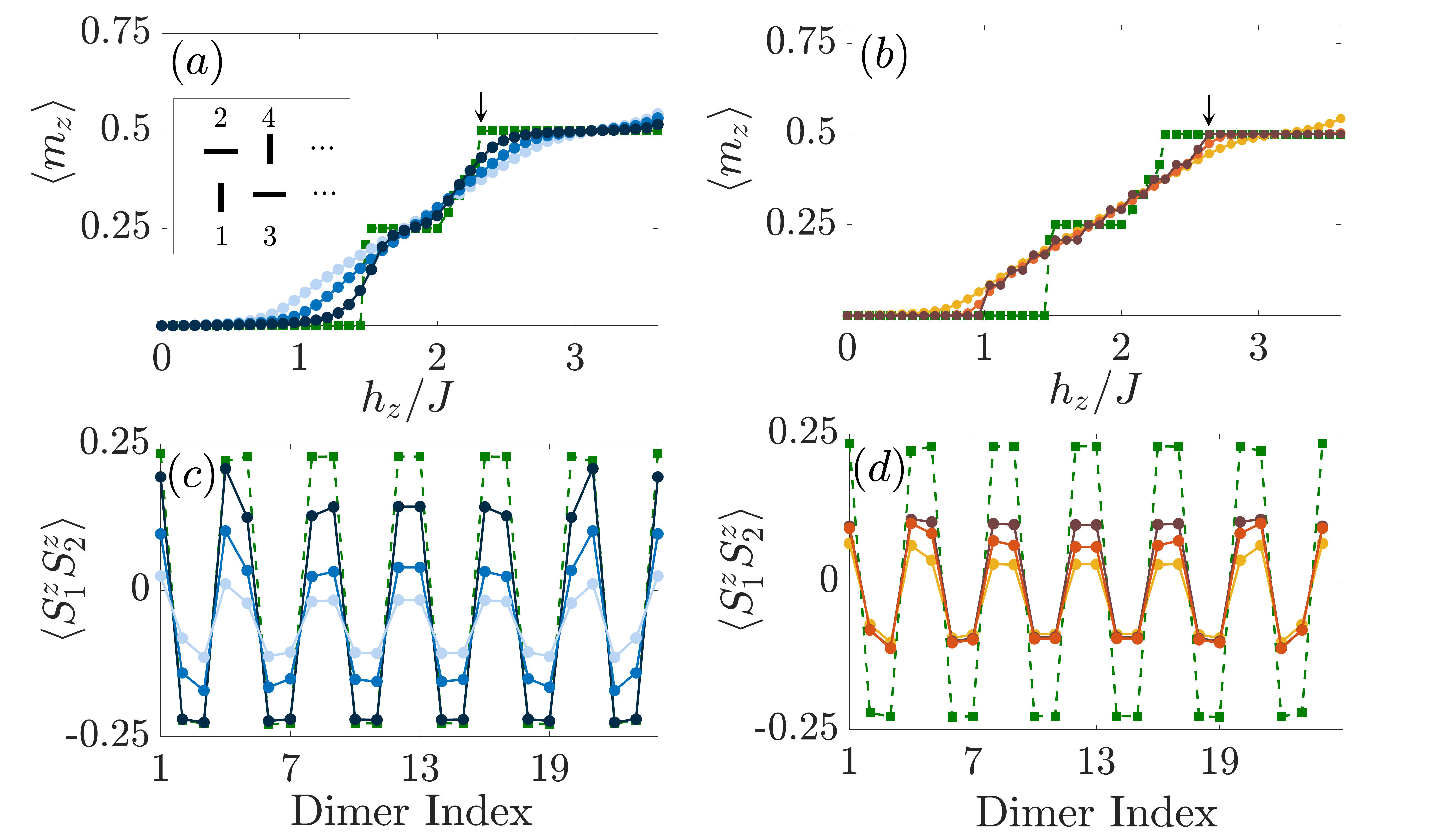}
\caption{\label{fig4}Characterization of dimer correlations for competing DM interactions: (a) magnetization plateau for no DM interactions (green dashed line with squares) and constant intradimer DM interactions $D/J=0.1$ with ascending interdimer DM interactions $D'_{\perp}/J= 0.00, 0.12, 0.20$ (darker circled blue line for smaller value). (Inset) Dimer index labeling conventions; (b) magnetization plateau for no DM interactions (green dashed line with circles) and constant interdimer DM interactions $D'_{\perp}/J=0.2$ with ascending intradimer DM interactions $D/J= 0.00, 0.02, 0.10$ (darker circled orange line for smaller value); (c) two-point correlation function $\langle S_1^{z}S_2^{z} \rangle$ on each dimer of the system for $h_z/J=2.32$ in (a) (indicated by black arrow), and (d) for $h_z/J=2.64$ in (b) (indicated by black arrow as well). The color of the lines in panels (c), (d) corresponds to those in panels (a), (b) respectively. Here, the subscripts $1$ and $2$ are the same site index in $(r,1)$ and $(r,2)$ on each dimer as in Fig. \ref{fig1_lattice}(a).} 
\end{figure*}

\subsection{Effect of competing DM interactions}
We now explore how the properties of the system are affected when both types of DM interactions are turned on. In Fig. \ref{fig4} we show the magnetization plateux and their corresponding two-point correlation functions $\langle S_i^z S_j^z \rangle$ on each dimer within the $1/2$ plateau as an example. 

First we fix the intradimer DM interactions at $D/J=0.1$ and we vary the interdimer DM interactions from $D'_{\perp}=0$ to $0.20$. We plot the magnetization curve in Fig. \ref{fig4}(a) and the correlation function $\langle S_i^z S_j^z \rangle$ at $h_z/J=2.32$ in Fig. \ref{fig4}(c). Each dimer is labeled in the order shown in the inset of Fig. \ref{fig4}(a). When there are no DM interactions, for the $1/2$ plateau the correlation has a pattern of one singlet followed by a triplet dimer within each unit cell as shown by the green dashed line. In Figs. \ref{fig4}(b), and \ref{fig4}(d) we fix the interdimer DM interactions at $D'_{\perp}/J=0.2$ and gradually increase the intradimer DM interactions from $D/J=0$ to $0.10$. Here the intradimer interaction removes the small plateaux because it breaks the number conservation. 

In the presence of intradimer DM interaction the $\langle S_i^z S_j^z \rangle$ are weaker in magnitude, and, as we increase the interdimer DM interactions, we observe a further decrease of the magnitude (darker lines correspond to smaller magnitudes of the interdimer DM interaction $D'_{\perp}$). 
We observe the singlet dimers are more robust against intradimer DM interactions, evidenced by the fact that the $\langle S_i^z S_j^z \rangle$ is not significantly affected when a weak intradimer DM interaction is turned on for the singlet dimers, while the triplet dimers are much more strongly affected. This is something that occurs also when analyzing small clusters with exact diagonalization \cite{edsmall}. 
The interdimer interaction instead, affects strongly both the singlet and the triplet dimers, as shown in Figs. \ref{fig4}(c), and \ref{fig4}(d). We remark that the fact that the intradimer interaction seems not to affect significantly the singlet dimers is accidental, as for different values of the magnetic field $h_z$ this does not occur.

\begin{figure}[ht]
\includegraphics[width=\columnwidth,draft=false]{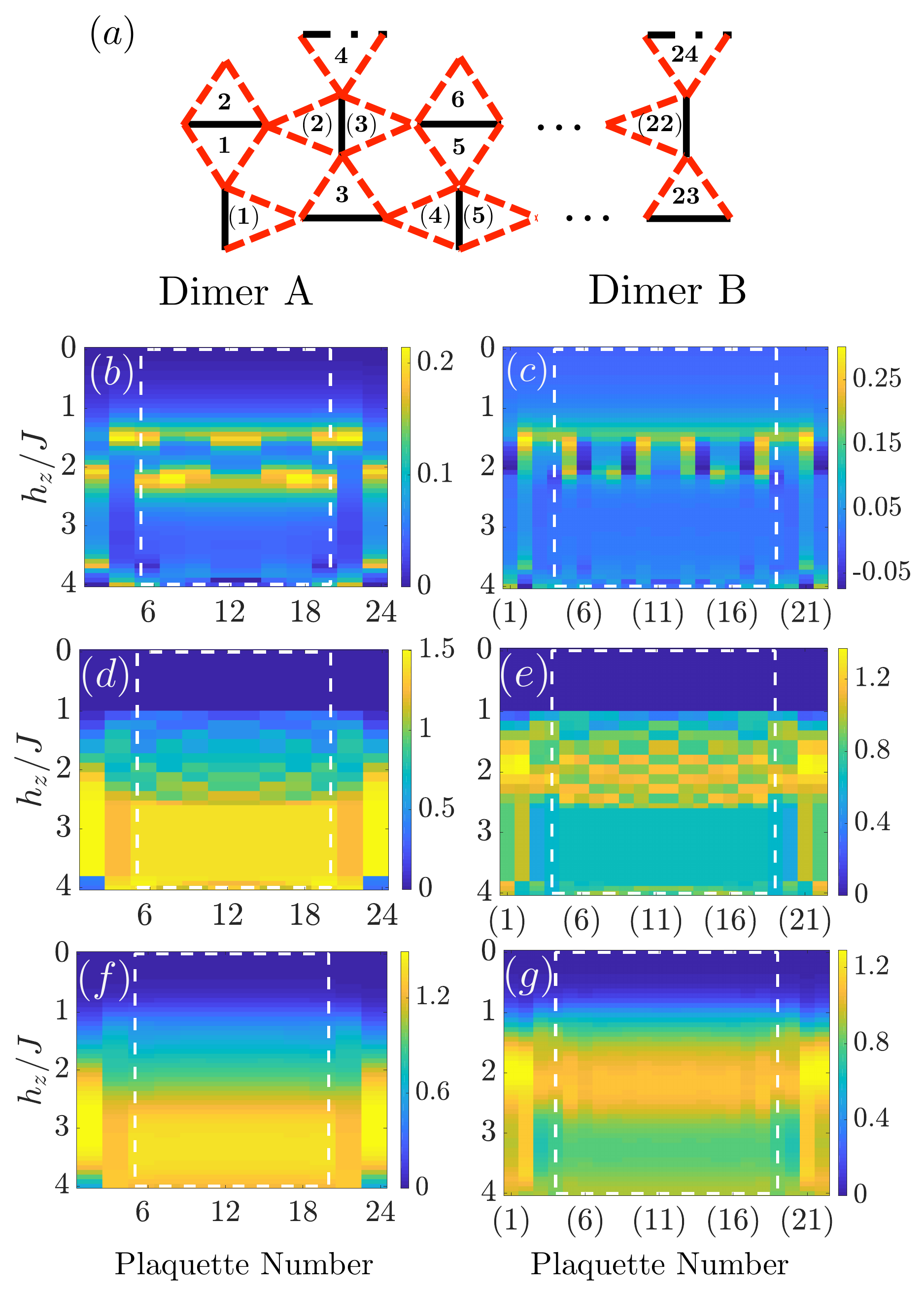}
\caption{\label{fig5}(a) Scalar spin chirality index convention on each triangle plaquette. The number with (without) parentheses corresponds to the plaquette number with a base on a $B$ ($A$) dimer; (b)-(g) Magnitude of scalar spin chirality associated to dimer $A$ (b),(d),(f) and $B$ (c),(e),(g) with respect to different types of DM interactions. The $x$ axis indicates the plaquette number, while the $y$ axis indicates the magnetic field $h_z$: (b),(c) with only intradimer DM interactions $D/J=0.1$, (d),(e) with only interdimer DM interactions $D'_{\perp}/J=0.2$, and (f),(g) with both DM interactions being present. The values corresponding to the middle eight rungs are highlighted by the white dashed squares.}
\end{figure}

\subsection{DM interaction induced chiral spin texture}
To better characterize the ground state of the system for different magnitudes of the magnetic fields and of the DM interactions, we also study the spin chirality{, a spin texture that is expected to emerge with such Hamiltonian terms}. We consider the triangular plaquettes associated respectively to the $A$ or $B$ dimers. The plaquettes are indexed following Fig. \ref{fig5}(a), from which, we highlight that there are $24$ plaquettes with a base on an $A$ dimer and $22$ plaquettes with a base on a $B$ dimer, listed with a number between parentheses $(\dots)$.  	 
The expression for the chirality \cite{KalmeyerLaughlin1987,WenWilczekZee1989} is given by 
\begin{align}
	\chi =8\bm{S_i} \cdot \left(\bm{S_j} \times \bm{S_k} \right)
\end{align} 
{The spin chirality is a quantitative measure of chiral order. It is zero for states such as collinear or coplanar magnetic states, for instance, ferromagnetic (FM), antiferromagnetic (AFM) while it is nonzero for non-coplanar spin configurations.} 
In Figs. \ref{fig5}(b)- \ref{fig5}(g) we show the magnitude of the chirality versus plaquette number for the $A$ dimers' plaquettes, Figs. \ref{fig5}(b), \ref{fig5}(d), and \ref{fig5}(f), and for the $B$ dimers' plaquettes, Figs. \ref{fig5}(c), \ref{fig5}(e), and \ref{fig5}(g). In particular, in Figs. \ref{fig5}(b), and \ref{fig5}(c) we consider only the effect of the intradimer DM interaction, in Figs. \ref{fig5}(d), and \ref{fig5}(e) we observe the effect of only interdimer DM interaction and in Figs. \ref{fig5}(f), and \ref{fig5}(g) the combined effect of both DM interactions. 
{In Fig. \ref{fig5}(b), and \ref{fig5}(c) we observe that chirality occurs primarily for values of the magnetic field corresponding to states between two plateaux, which are regions of the parameter space which are less robust to DM interaction. In Fig. \ref{fig5}(d), and \ref{fig5}(e), when it is only in the presence of interdimer DM interactions, } 
chirality emerges already for magnetic fields smaller than required to form the $1/4$ plateau. In this case, the response in the $A$ and $B$ dimers is very different, with the $A$ dimers displaying an overall increasing chirality in the plaquettes from $h_z=0$ to the remnants of the $1/2$ plateau, while in the $B$ dimers the chirality is maximum for fields corresponding to the $1/4$ plateau. 
{For larger interdimer interaction $D'_{\perp}/J=0.2$, together with a finite intradimer interaction $D/J=0.1$, the response is smoother versus the plaquette number, as shown in Figs. \ref{fig5}(f), and \ref{fig5}(g).}       

\begin{figure}
\includegraphics[width=\columnwidth,draft=false]{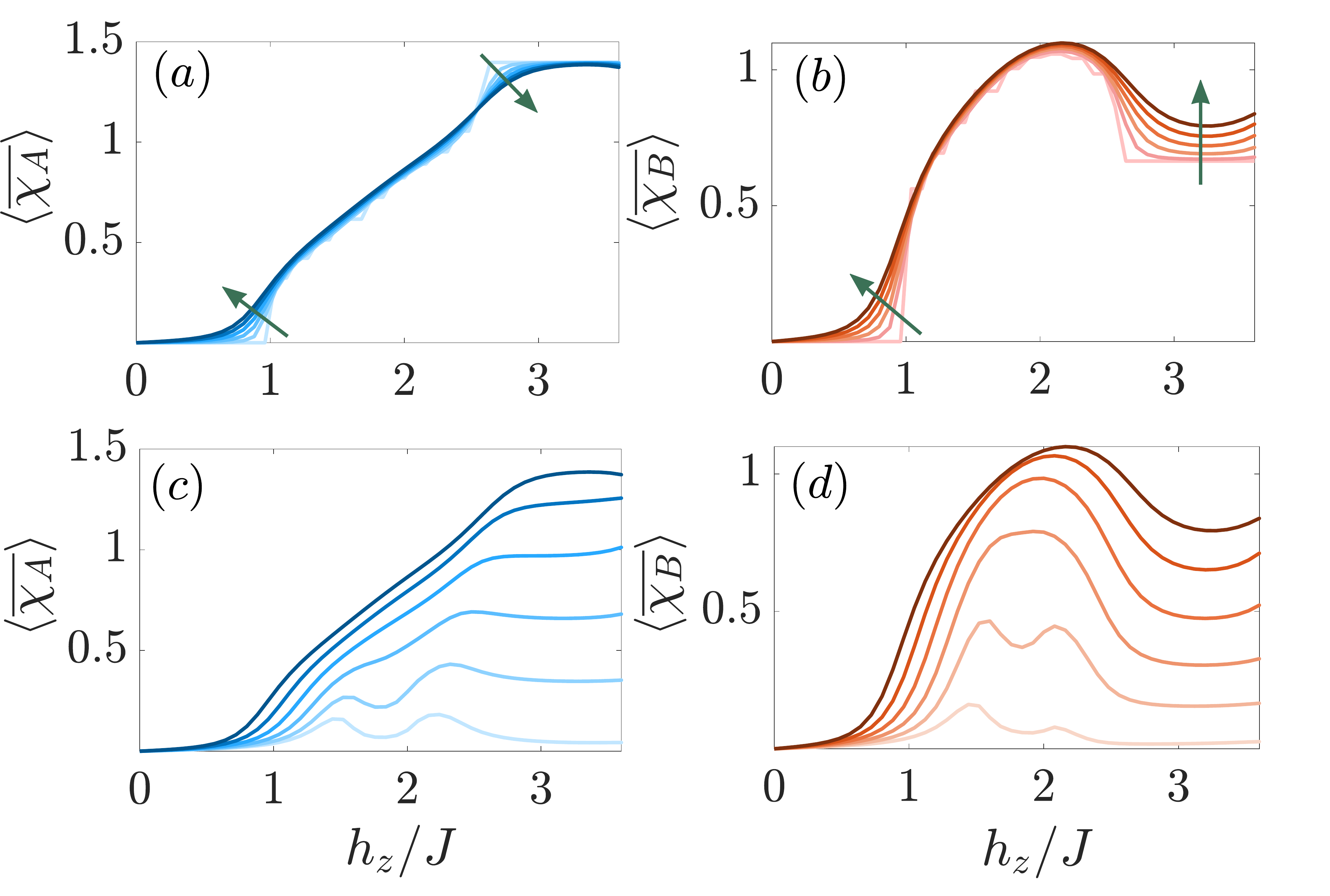}
\caption{\label{fig6} Averaged scalar spin chirality in the bulk as a function of magnetic field $h_z$ for dimers A and B: (a) averaged scalar spin chirality for dimer $A$ and (b) for dimer $B$. The interdimer DM interactions $D'_{\perp}/J=0.2$ and the intradimer DM interactions increase from $D/J=0.0$ to $0.1$ with step $0.02$ (from lighter to darker color and indicated by an arrow as well). (c) Averaged scalar spin chirality for dimer $A$ and (d) for dimer $B$. The intradimer DM interactions $D/J=0.1$ and the interdimer DM interactions increase from $D'_{\perp}/J=0.0$ to $0.2$ with step $0.04$ (also  from lighter to darker color).}
\end{figure}
In Fig. \ref{fig6} we show the average value of the chirality in the bulk of the system, in panels (a), and (c) for the $A$ dimers and (b), and (d) for the $B$ dimers. In Figs. \ref{fig6}(a), and \ref{fig6}(b) we consider the case of a small interdimer DM interaction $D'_{\perp}/J=0.2$ and then we perturb it with intradimer DM interaction with values $D/J$ from $0$ to $0.1$. 
{We observe that, dimer $A$ and dimer $B$ have distinct response to the intradimer DM interaction, which is due to the narrow geometry of the system: the chirality corresponding to dimer A monotonically increases for all range of the magnetic field, while the chirality corresponding to dimer B actually decreases within the magnetic field with respect to the $1/2$ plateau [indicated by arrows in Figs. \ref{fig6}(a), and \ref{fig6}(b)]. }
As for magnetization plots, the intradimer interaction makes the curves smoother. In Figs. \ref{fig6}(c), and \ref{fig6}(d) we depict the case in which the intradimer interaction is fixed at $D/J=0.1$ and we increase the interdimer DM interaction from $D'_{\perp}=0$ to $0.2$. 
{Consistently with the previous discussion of Figs. \ref{fig5}(b), and \ref{fig6}(c)}, here we observe that in the absence of interdimer DM interaction the chirality is only present between the plateaux, and, as $D'_{\perp}$ increases, chirality increases significantly across the values of magnetic field observed. We stress that the combination of both interdimer and intradimer DM interactions results in a homogeneous chirality within the plaquettes of the $A$ or $B$ dimers{ [see also Figs. \ref{fig5}(f), and \ref{fig5}(g)]}.

{Moreover, the distinct effect of intradimer and interdimer DM interaction on the chirality can be identified through observing the lightest colored curves in Figs. \ref{fig6}(c), and \ref{fig6}(d): the interdimer DM interaction is responsible for the emergence of noncoplanar configuration both between and within the magnetic field range of plateaux [Figs. \ref{fig6}(a), and \ref{fig6}(b)] while the intradimer DM interaction tend to primarily induce the noncoplanar spin configuration only between the plateaux [Figs. \ref{fig6}(c), and \ref{fig6}(d)]. In general, the interdimer DM interaction changes the average value of chirality more significantly than the intradimer DM interactions. This is mainly due to the fact that we consider the case with $J’/J=0.5$ for which the dimers are more strongly coupled.}                              

\section{\label{sec:conclusion}Conclusion and Discussion}
In summary, we have investigated a quasi-two-dimensional Shastry-Sutherland model with competing non-uniform Dzyaloshinskii-Moriya interactions and also an external magnetic field in the $z$ direction. We have used a matrix product state algorithm and numerically obtained the magnetization curve with and without DM interactions. 
{We have shown that both types of DM interactions affect significantly the plateaux, destroying them. The intradimer DM interaction clearly smoothens the magnetization curve, while the interdimer DM interaction analyzed, since it is number-conserving, would generate a smooth curve only in the thermodynamics limit. Furthermore, we have investigated the effect of intradimer and interdimer DM interactions by probing with different two-point correlation functions.} We have also explored the DM interaction induced chiral spin texture from the observation of scalar spin chirality on triangular plaquettes {showing that intradimer DM interactions induce chirality more easily for states between magnetization plateaux, while interdimer DM interaction can induce it also in regions in which the magnetization is flat as a function of the magnetic field.}  {For different geometries or values of $J'/J$ it would be possible to study the effect of DM interactions on different plateaux, in which the origin of the plateau is due to different spin structures like, e.g., for the $1/8$ plateau \cite{CorbozMila2014,FoltinManmanaSchmidt2014}. For this we would have to analyze longer systems or even two-dimensional ones, something that it is currently beyond the techniques we employ. }    

Given the effect of DM interactions on the magnetization plateaux and on the emergence of chirality, in future works we will focus on how the DM interactions affect the transport properties of the system. Another interesting research direction is to consider how these phases are robust to different dissipative perturbations, both for their relaxation dynamics and the steady state reached.    

\begin{acknowledgments}
We acknowledge A. Honecker for fruitful discussions and for feedback on the manuscript. D.P. and P.S. acknowledge support from the Singapore Ministry of Education, Singapore Academic Research Fund Tier-II (Project No. MOE2016-T2-1-065). Our numerical simulation was partially conducted on resources of National Supercomputing Centre, Singapore (NSCC)\cite{NSCC}.
\end{acknowledgments}

\bibliographystyle{apsrev}

\begin{thebibliography}{100}
\bibitem{Diep2013} H.T. Diep, {\it Frustrated Spin Systems}, World Scientific (2013).   
\bibitem{ShastrySutherland1981} B. S. Shastry and B. Sutherland, Physica B {\bf 108}B, 1069 (1981). 
\bibitem{MisguichGirvin2001} G. Misguich, Th. Jolicoeur, and S.M. Girvin, Phys. Rev. Lett. {\bf 87}, 097203 (2001).     
\bibitem{DorierMila2008} Dorier J., Schmidt K. P. and Mila F., Phys. Rev. Lett. {\bf 101}, 250402 (2008).    
\bibitem{AbendscheinCapponi2008} Abendschein A. and Capponi S., Phys. Rev. Lett. {\bf 101}, 227201 (2008).   
\bibitem{ManmanaMila2011} S. R. Manmana, J.-D. Picon, K.P. Schmidt, and F. Mila, EuroPhys. Lett. {\bf 94}, 67004 (2011).    
\bibitem{JaimeGoulin2012} Marcelo Jaime, Ramzy Daou, Scott A. Crooker, Franziska Weickert, Atsuko Uchida, Adrian E. Feiguin, Cristian D. Batista, Hanna A. Dabkowska, and Bruce D. Gaulin, Proc. Nat. Acad. Sc. {\bf 109}, 12404 (2012). 
\bibitem{NemecSchmidt2012} M. Nemec, G.R. Foltin, and K.P. Schmidt, Phys. Rev. B {\bf 86}, 174425 (2012). 
\bibitem{MatsudaMila2013} Y.H. Matsuda, N. Abe, S. Takeyama, H. Kageyama, P. Corboz, A. Honecker, S.R. Manmana, G.R. Foltin, K.P. Schmidt, and F. Mila, Phys. Rev. Lett. {\bf 111}, 137204 (2013).    
\bibitem{CorbozMila2014} P. Corboz, and F. Mila, Phys. Rev. Lett. {\bf 112}, 147203 (2014).    
\bibitem{WietekHonecker2019} A. Wietek, P. Corboz, S. Wessel, B. Normand, F. Mila, and A. Honecker, Phys. Rev. Research {\bf 1}, 033038 (2019).              
\bibitem{KageyamaUeda1999} H. Kageyama, K. Yoshimura, R. Stern, N. V. Mushnikov, K. Onizuka, M. Kato, K. Kosuge, C. P. Slichter, T. Goto, and Y. Ueda, Phys. Rev. Lett. {\bf 82}, 3168 (1999).    
\bibitem{OnizukaGoto2000} K. Onizuka, H. Kageyama, Y. Narumi, K. Kindo, Y. Ueda, T. Goto, J. Phys. Soc. Jpn. {\bf 69}, 1016 (2000).  
\bibitem{KodamaMila2002} K. Kodama, M. Takigawa, M. Horvati\'{c}, C. Berthier, H. Kageyama, Y. Ueda, S. Miyahara, F. Becca, F. Mila, Science {\bf 298}, 395 (2002).
\bibitem{RichterSchulenburg1998} J. Richter, N.B. Ivanov, and J. Schulenburg, J. Phys.: Condens. Matter {\bf 10}, 3635 (1998).   
\bibitem{KogaKawakami2000} A. Koga, K. Okunishi, and N. Kawakami, Phys. Rev. B {\bf 62}, 5558 (2000).   
\bibitem{SchulenburgRichter2002} J. Schulenburg and J. Richter, Phys. Rev. B {\bf 65}, 054420 (2002).     
\bibitem{HoneckerRichter2004} A. Honecker, J. Schulenburg, and J. Richter, J. Phys.: Condens. Matter {\bf 16}, S749 (2004).     
\bibitem{OhanyanHoenecker2012} V. Ohanyan, and A. Honecker, Phys. Rev. B {\bf 86}, 054412 (2012).   

\bibitem{VerkholyakStrecka2013} T. Verkholyak, and J. Stre\v{c}ka, Phys. Rev. B {\bf 88}, 134419 (2013).  
\bibitem{BarredoBrowayes2016} D. Barredo, S. de L\'{e}s\'{e}leuc, V. Lienhard, T. Lahaye, and A. Browaeys, Science {\bf 354}, 1021 (2016).
\bibitem{EndresLukin2016} M. Endres, H. Bernien, A. Keesling, H. Levine, E.R. Anschuetz, A. Krajenbrink, C. Senko, V. Vuletic, M. Greiner, and M. D. Lukin, Science {\bf 354}, 1024 (2016).    
\bibitem{RoomKageyama2004} T. R\~{o}\~{o}m, D. H\"{u}vonen, U. Nagel, J. Hwang, T. Timusk, H. Kageyama, Phys. Rev. B {\bf 70}, 144417 (2004).   
\bibitem{LevyUeda2007} F. Levy, I. Sheikin, C. Berthier, M. Horvati\'{c}, M. Takigawa, H. Kageyama, T. Waki and Y. Ueda, EuroPhys. Lett. {\bf 81}, 67004 (2007).       
\bibitem{DalibardOhberg2011} J. Dalibard, F. Gerbier, G. Juzeli\={u}nas, and P. \"{O}hberg, Rev. Mod. Phys. {\bf 83}, 1523 (2011). 
\bibitem{GoldmanSpielman2014} N. Goldman, G. Juzeliunas, P. Ohberg, I. B. Spielman, Rep. Prog. Phys. {\bf 77} 126401 (2014).   
\bibitem{RomhanyiPenc2011} J. Romh\'{a}nyi, K. Totsuka, and K. Penc, Phys. Rev. B {\bf 83}, 024413 (2011).     
\bibitem{fn1} This interaction is expected to be present at low temperature\cite{RomhanyiPenc2011}.      
\bibitem{fn2} This form of interaction is for the high symmetry case, which occurs at high enough temperature\cite{RomhanyiPenc2011}.     
\bibitem{Schollwock2011} U. Schollw\"ock, Ann. Phys. {\bf 326}, 96 (2011).  
\bibitem{White1992} S.R. White, Phys. Rev. Lett. {\bf 69}, 2863 (1992).     
\bibitem{StoudenmireWhite2012} E.M. Stoudenmire, S.R. White, Ann. Rev. Cond. Matt. Phys. {\bf 3}, 111 (2012).    
\bibitem{White2005} S.R. White, Phys. Rev. B {\bf 72}, 180403 (2005). 
\bibitem{edsmall} For instance, this has already been observed with a system of only eight spins in total.  
\bibitem{KalmeyerLaughlin1987} V. Kalmeyer and R. B. Laughlin, Phys. Rev. Lett. {\bf 59}, 2095 (1987). 
\bibitem{WenWilczekZee1989} X. G. Wen, F. Wilczek, and A. Zee, Phys. Rev. B {\bf 39}, 11413 (1989).
\bibitem{FoltinManmanaSchmidt2014} G. R. Foltin, S. R. Manmana, and K. P. Schmidt, Phys. Rev. B {\bf 90}, 104404 (2014). 
\bibitem{NSCC} \texttt{https://www.nscc.sg/}    

\end{thebibliography}

\end{document}